\documentclass[prb,twocolumn,showpacs]{revtex4}% Physical Review B
\usepackage{dcolumn}% Align table columns on decimal point
\usepackage[dvips]{graphicx}
\usepackage{latexsym}
\usepackage{amsmath}
\usepackage{amssymb}
\usepackage{bm}% bold math
\begin{document}
%%%%%%%%%%%%%%%%%%%%%%%%%%%%
\newcommand{\be}{\begin{equation}}
\newcommand{\ee}{\end{equation}}
\newcommand{\ba}{\begin{eqnarray}}
\newcommand{\ea}{\end{eqnarray}}
\newcommand{\la}{\langle}
\newcommand{\ra}{\rangle}
\newcommand{\vk}{{\bf k}}
\newcommand{\vq}{{\bf q}}
\newcommand{\vp}{{\bf p}}
\newcommand{\vx}{{\bf x}}
\newcommand{\half}{\frac{1}{2}}
%%%%%%%%%%%%%%%%%%%%%%%%%%%%%%%%%%%%%%%%%%%%%%%%%%%%%%%%%%%%%%%%%
\title{ Slave-boson approach to  the infinite-$U$  Anderson-Holstein impurity model}
\author{Hyun C. Lee}
\email{hyunlee@sogang.ac.kr}
\affiliation{Department of Physics and Basic Science Research Institute,
Sogang University, Seoul, 121-742, Korea}
\author{Han-Yong Choi}
\affiliation{Department of Physics,
Department of Physics, BK21 Physics Research Division, and Institute of Basic Science Research, \\
Sung Kyun Kwan University, Suwon, 440-746, Korea.}
\date{\today}
%%%%%%%%%%%%%%%%%%%%%%%%%%%%%%%%%%%%%%%%%%%%%%%%%%%%%%%%
\begin{abstract}
The infinite-$U$ Anderson-Holstein impurity model is studied with a focus
on the interplay between the strong electron correlation and the weak electron-phonon interaction.
The slave boson method has been employed in combination with the large
degeneracy expansion ($1/N$) technique. The charge and spin susceptibilities and the phonon propagator are obtained in the
approximation scheme where the saddle point configuration and the
Gaussian $1/N$  fluctuations are taken into account. The spin
susceptibility is found not to be renormalized by electron-phonon
interaction, while the charge susceptibility is renormalized.
From the renormalized charge susceptibility the Kondo temperature
is found to increase by the electron-phonon interaction. It turns
out that the bosonic $1/N$ Gaussian fluctuations play a very
crucial role, in particular, for the phonon propagator.
\end{abstract}
%%%%%%%%%%%%%%%%%%%%%%%%%%%%%%%%%%%%%%%%%%%%%%%%%%%%%%%%%%%
\maketitle

%%%%%%%%%%%%%%%%%%%%%%%%%%%%%%%%%%%%%%%%%%%%%%%%%%%%%%%%%%%%
\section{Introduction}
Both electron correlation and lattice dynamics are important in
understanding the physical properties of many condensed matter
systems.\cite{orbital} But the {\it interplay} between them has
not been studied in detail because there are no reliable
theoretical methods to describe the electron-phonon interaction
in the strongly correlated electron systems. For weakly correlated
materials we have the successful Migdal-Eliashberg
theory.\cite{ME} There is, however, no systematic theoretical
extension of the theory to the strongly interacting systems.
With the recent advent of the dynamical mean field theory (DMFT)
\cite{dmft} various lattice problems can be mapped (or
approximated) onto
impurity problems. In this context it is
important to understand the correlated impurity problem coupled
to local phonons.

One of the simplest model which incorporate  both electron
correlation and phonon is the Hubbard-Holstein (HH) model.\cite{jeon,hewson2} In the context of DMFT,
it is mapped onto the Anderson-Holstein (AH)
model.\cite{hewson,jpc} The AH model is a single-impurity Anderson
model with a linear coupling to a local phonon mode as in the
Holstein model.\cite{holstein} Recently, the HH model was studied with DMFT in combination with the numerical
renormalization group (NRG) at half-filling at zero temperature.\cite{jeon,hewson3} 
The NRG method has also been successfully applied
to AH model\cite{hewson,jpc}, and almost exact results on the
electron and phonon spectral functions have been obtained.
However, it is still desirable  to develop an analytic scheme in
spite of its approximate nature since it helps us  to understand
the underlying physics in more clear and intuitive way.
In this paper, we will consider the regime where the electron correlation dominates 
over the phonon effect, which will allow us to study the phonon effects on the Kondo physics.
The pure Holstein model without electron correlation is studied
in Ref. \onlinecite{zeyher} using semiclassical approximation 
and in Ref. \onlinecite{holstein2} using NRG and DMFT.

When the local Coulomb repulsion of Anderson model is very large
compared to the hybridization between the impurity and the
conduction electrons, the impurity develops a magnetic moment at
high temperature regime.\cite{hewson2} The moment is gradually
screened by the spin of the conduction electrons as the
temperature is lowered below the Kondo temperature scale. The
physical properties of Anderson model at temperature lower than
the Kondo temperature can not be described within perturbative
framework since the \textit{effective} Kondo coupling constant
becomes very large in that temperature regime. This low temperature state is often called
the Kondo fixed point in the sense of renormalization
group.\cite{hewson2,comment} We mention that the physics of Kondo
fixed point can describe the correlated metallic states such as
heavy fermions in the context of DMFT.

In the limit of the {\it infinite} local Coulomb repulsion the
impurity  develops the local moment at high temperature and the
system crosses over into the (completely screened) Kondo fixed
point at low temperature. Thus the physical properties of the
system in the limit of {\it infinite} local Coulomb repulsion are
expected to be qualitatively  similar to those of other physical systems
with very large but  finite  local Coulomb repulsion. If the
local Coulomb repulsion is taken to be infinite the doubly
occupied state of the impurity is not allowed in the physical
Hilbert space. In general, the standard perturbative approach
based on the Wick theorem on the electron operators and the
associated Feynman diagram expansions does not apply to the
systems with the constrained Hilbert space. This fact poses a
technical difficulty for the approaches based on the
\textit{electron} operators. However, if we employ the slave
field representation of electron operators, the constrained
Hilbert space  can be treated in a considerably simplified way.
The slave field method also admits the Feynman diagram expansion
and the direct manipulations of  functional integrals. This
slave field method has been successfully applied to the
infinite-$U$ Anderson model\cite{coleman,read,read2} and it is particularly adequate in
the study of the low temperature properties of the Anderson
model. The bose condensed state which
emerges in the saddle point (mean-field) approximation
embodies the Kondo resonant state, which is the crux of the Kondo
fixed point. Thus the essential physics of the Kondo fixed point
is captured in the leading approximation of the slave boson
method. At high temperature the bose condensate disappears and it
corresponds to the high temperature \textit{weak} Kondo coupling regime
where the perturbative approach such as the poor man's
scaling\cite{haldane} can be applied.

Our objective in this paper is to incorporate the phonon dynamics
into the infinite-$U$ Anderson model within the scheme of the
slave boson  method along the line pursued by Coleman.\cite{coleman} We
will assume that the electron-phonon interaction is weak compared
to the strong electron correlation, so that the physical
situation at the zeroth approximation is still best described in
terms of Kondo fixed point physics. We have developed a scheme in
which the effect of electron-phonon interaction can  be computed
systematically in the large-degeneracy limit or, in other words,
the  $1/N$ expansion. The basic idea of the $1/N$ expansion is briefly
illustrated in Appendix \ref{1/N}.

The charge and spin susceptibilities of the impurity and the
local phonon propagator have been obtained. Note that the
dynamical correlation functions are generally very hard to compute
when the electron correlation is very strong. The feasibility of
the analytic computation of the dynamical correlation functions
is the great advantage of the slave boson methods over other methods.
It is found that the spin susceptibility of the impurity
[Eq.~(\ref{spinsusc1})] is not modified by the electron-phonon interaction
up to the approximation of
Gaussian fluctuations. The charge susceptibility of the impurity
up to the second order in the (weak) electron-phonon coupling is
given by Eq.~(\ref{charge2}). From the charge susceptibility at
zero frequency,  the Kondo temperature renormalized by the
electron-phonon interaction can be extracted, and it is given by
Eq.~(\ref{main2}). Finally the phonon propagator is found to be
Eq.~(\ref{phononprop1}). These are the main results of this paper.
In all of these results the influence of electron-phonon
interaction is small since the charge fluctuations which directly 
couple to the local phonon are severely suppressed by very strong electron correlations. 
These results manifestly demonstrate the essential \textit{interplay} between the strong electron 
correlation and the (weak) electron-phonon interaction. For instance, 
 the small renormalization by electron-phonon interaction in Eq.~(\ref{main2})
  is expressed by a \textit{product} of the polaron energy and the Kondo energy,
where the nonpeturbative physics of the electron correlation is encapsulated in the Kondo energy scale.
The smallness of the influence of the electron-phonon interaction in our case 
is partially due to the absence of the orbital
degrees of freedom.  The effect of phonons can be much more substantial for the materials with multi-orbitals,
where the phonons can couple to the orbital quantum number.\cite{orbital}  
Our approach can be extended to such cases, and the progresses in this direction are being made. 

This paper is organized as follows. In Sec. II we introduce the
AH model and the slave boson formalism. Then the rescaled
effective action in the Read-Newns gauge is obtained. In Sec. III
we summarize the basic results on the  saddle
point approximation by Coleman.\cite{coleman} In Sec. IV the
phonon dynamics is coupled to the bose excitation emerging in
the $1/N$ expansion and the effective action of them is computed.
In Sec. V the spin and the charge susceptibilities are calculated
from the effective action obtained in Sec. IV and their physical
properties are analyzed. In Sec. VI the phonon propagator is
calculated from the effective action obtained in Sec. IV. Sec.
VII is for the summary and concluding remarks. Some of details of
the calculations can be found in the Appendices.

%%%%%%%%%%%%%%%%%%%%%%%%%%%%%%%%%%%%%%%%%%%%%%%%%%%%%%%%%%%
\section{Formalism}

\subsection{Slave boson representation}
The Hamiltonian of AH model consists of three parts. \be {\cal
H}_{AH}={\cal H}_{el}+{\cal H}_{ph}+{\cal H}_{el-ph}. \ee ${\cal
H}_{el}$ and ${\cal H}_{ph}$ is the Hamiltonian for electron and
phonon part, respectively, and ${\cal H}_{el-ph}$ is the
Hamiltonian describing the interaction between the electron and
the phonon. \ba {\cal H}_{el}&=&\sum_{k,\sigma} \epsilon_k
\,c^\dag_{k\sigma} c_{k \sigma}
+ \epsilon_f \sum_\sigma f^\dag_\sigma f_\sigma + U f^\dag_\uparrow f_\uparrow  f^\dag_\downarrow f_\downarrow \nonumber \\
&+& \sum_{k \sigma} \frac{1}{\sqrt{N_{lat}}} ( V_k f^\dag_\sigma c_{k \sigma}+ V_k^*c^\dag_{k \sigma} f_\sigma).
\ea
\be
{\cal H}_{ph}= \frac{1}{2}\Big( M \Omega^2 Q^2+ \frac{P^2}{M} \Big).
\ee
\be
{\cal H}_{el-ph}= g_0 Q \Big(  \sum_\sigma f^\dag_\sigma f_\sigma -\langle \sum_\sigma f^\dag_\sigma f_\sigma \rangle \Big).
\ee
$\sigma=\uparrow,\downarrow$ is the spin index, and $V_k$ is the hybridization matrix element.
$f_\sigma$ is the impurity electron operator, and $c_{k \sigma}$ is the conduction electron operator.
$\epsilon_f$ is the energy of the impurity level, and $U$ is the local Coulomb repulsion at the impurity site.
$Q$ is local phonon coordinate. $P$ is the conjugate momentum of $Q$ satisfying
$[Q,P]=i \hbar$.
$M$ is ion mass and $\Omega$ is the oscillator frequency of dispersionless (Einstein) phonons.
$N_{lat}$ is the number of lattice sites for the conduction electrons.
$g_0$ is the (unscaled) electron-phonon coupling constant.

In the limit of infinite $U$, the doubly occupied state of
impurity  $f$ electrons is prohibited. In this {\it constrained}
Hilbert space, the impurity electron operator $f_\sigma$ can be
expressed in terms of fermion operator $s_\sigma$ and the slave
boson operator $b$ satisfying\cite{coleman,notation} \ba
\label{slave}
& &f^{\dag}_\sigma= s^\dag_\sigma  b ,\;\; f_\sigma = s_\sigma b^\dag,  \\
\label{constraint}
& &b^\dag b + \sum_\sigma s^\dag_\sigma s_\sigma=1.
\ea
Now the remaining empty state and the singly occupied states denoted by spin quantum number $\sigma$
 at the impurity site can be represented as follows.
\ba
|\textrm{empty} \rangle &=& b^\dag | \textrm{ref} \rangle, \nonumber \\
|\sigma \rangle &=& s^\dag_\sigma | \textrm{ref} \rangle,
\ea
where $ | \textrm{ref} \rangle$ denotes some arbitrary \textit{reference} state.
The local Coulomb repulsion term $U f^\dag_\uparrow f_\uparrow  f^\dag_\downarrow f_\downarrow$
which is non-zero only for the doubly occupied states can be dropped in the constrained Hilbert space.
{\it After} the elimination of the doubly occupied state we can formally increase
the number of spin components from 2  to arbitrary $N$.

The  Hamiltonian in the constrained Hilbert space can be expressed in terms of
the slave fields as follows.
(the index $\sigma$ has been changed to $m$ in anticipation of $1/N$ expansion)
\ba
\label{qqqq}
H&=&\sum_{k,m=1}^N (\epsilon_k - \mu_c)\,c^\dag_{k m} c_{k m}
+ \epsilon_f \sum_m s^\dag_m s_m \nonumber \\
&+& \frac{1}{\sqrt{N_{ lat}}}\,\sqrt{\frac{1}{N}}\,\sum_{k m}\,
(V_k s^\dag_m b c_{k m} + V_k^*  c_{k m}^\dag b^\dag s_m) \nonumber \\
&+&\frac{1}{2}\Big( M \Omega^2 Q^2+ \frac{P^2}{M} \Big) \nonumber \\
&+&g_0 Q (  \sum_m s^\dag_m s_m  -\langle \sum_m s^\dag_m s_m  \rangle).
\ea
The electron-phonon coupling gives rise to the {\it charge} fluctuation since it couples to
the {\it total} charge at the impurity site which is either 0 or 1 in the constrained Hilbert space.
Note the scaling of the hybridization matrix element $V_k \to \sqrt{\frac{1}{N}} V_k$ in Eq.~(\ref{qqqq}).
$\mu_c$ is the chemical potential for the conduction electrons.
At this point we pass into the Lagrangian formulation in imaginary time.
\ba
L&=&\sum_{ km } c^{\dag}_{km} \Big[ \partial_\tau +\epsilon_k - \mu_c \Big ]\,c_{km}  \nonumber \\
&+&\sum_m s_m^{\dag} \Big[ \partial_\tau  + \epsilon_f \Big] s_m
 +b^\dag \partial_\tau b \nonumber \\
 &+& \frac{1}{\sqrt{N_{{\rm lat}}}}\,\sqrt{\frac{1}{N}}\,\sum_{k m}\,
(V_k s^\dag_m b c_{k m} + V_k^*  c_{k m}^\dag b^\dag s_m) \nonumber \\
&+&\frac{M}{2}\Big((\partial_\tau Q)^2+  \Omega^2 Q^2 \Big) \nonumber \\
&+&g_0 Q \Big(  \sum_m s^\dag_m s_m - \langle \sum_m s^\dag_m s_m \rangle \Big)\nonumber \\
&+&\lambda( \sum_m s_m^{\dag} s_m + b^\dag b -1),
\ea
where $\lambda$ is a Lagrange multiplier implementing the constraint Eq.~(\ref{constraint}).
In the $1/N$ expansion this constraint will be realized through
 the conserving property of the $1/N$ expansion.\cite{coleman}
The partition function is given by the following functional integral
\be
Z=\int D[b,\lambda] \int D[s,Q] \int D[c] \,e^{-\int d \tau L}.
\ee

The integration over the conduction electrons $c_{km}$ can be done exactly and it generates
 a self-energy $\Sigma_0(i \epsilon)$ for the slave fields.
\ba
\label{eq:self0}
\Sigma_0(i\epsilon)&=&\frac{1}{N_{lat}}\,\sum_k\,\frac{|V_k|^2}{i\epsilon+\mu - \epsilon_k} \nonumber \\
&\approx& {\rm Re} \Sigma_0 - i \Delta_0\,{\rm sgn}(\epsilon),\nonumber \\
\Delta_0&=&\pi N(E_F)\,\langle |V_k|^2 \rangle_{FS},
\ea
where $N(E_F)$ is the density of states of the conduction electrons
at Fermi energy and the subscript $FS$ denotes an average
over Fermi surface. The hybridization matrix element $V_k$ was assumed to be
weakly dependent on momentum near the Fermi surface of conduction electrons in the second line
of Eq.~(\ref{eq:self0}).
From now on ${\rm Re} \Sigma_0$ will be neglected since it is the order of the inverse of energy cutoff.
After the integration over the conduction electrons $c_{km}$, the resulting action becomes
\ba
S&=& \int_0^\beta d\tau \,\Big[ \sum_m s_m^{\dag} ( \partial_\tau  + \epsilon_f ) s_m
 +b^\dag \partial_\tau b \Big ] \nonumber \\
 &+& \frac{1}{N}\,\int_0^\beta d \tau d \tau^\prime \sum_m \Sigma_0(\tau-\tau^\prime)\,
 s_m^{\dag}(\tau) b(\tau)  s_m(\tau^\prime) b^\dag(\tau^\prime) \nonumber \\
 &+&\int_0^\beta d\tau \frac{M}{2}\Big((\partial_\tau Q)^2+  \Omega^2 Q^2 \Big)  \nonumber \\
 &+&\int_0^\beta d\tau \,g_0 Q \,
 \Big(  \sum_m s^\dag_m s_m -  \langle \sum_m s^\dag_m s_m \rangle \Big )\nonumber \\
 &+&\int_0^\beta d\tau\, \lambda\, ( \sum_m s_m^{\dag} s_m + b^\dag b -1),
 \ea
where
\be
\Sigma_0(\tau-\tau^\prime)=T \sum_{i\epsilon} e^{-i\epsilon(\tau-\tau^\prime)}\,
\Sigma_0(i\epsilon).
\ee
%%%%%%%%%%%%%%%%%%%%%%%%%%%%%%%%%%%%%%%%%%%%%
\subsection{Read-Newns gauge and Rescalings}
The systematic computations of $1/N$ fluctuations can be most clearly achieved in the
Read-Newns (RN) gauge:\cite{coleman,read,read2}
\be
\label{def:rn}
b(\tau)=R(\tau)\,e^{i \theta(\tau)},\;\;
s_m(\tau)=z_m(\tau)\,e^{i \theta(\tau)}.
\ee
Note that the original $f$-electron operator is invariant under the following gauge transformation:
\be
s_m  \to  s_m  e^{i \varphi},\;\;
b  \to b e^{ i \varphi}.
\ee
$R(\tau) \ge 0$ is the modulus of complex boson $b$.
Following Coleman \cite{coleman} we write the fluctuating Lagrange multiplier as
\be
\lambda = i \Omega_0 + \lambda_{sa},
\ee
where $\lambda_{sa}$ is the real part of the saddle point value of the Lagrange multiplier, which is
to be determined later.
In RN gauge, the infrared divergences which plague  the expansion in terms of the
original fields $b, s_m$ can be avoided.

The action in RN gauge is expressed as follows:
\ba
\label{action:rn}
& &S=\int_0^\beta d \tau \,
\sum_m z_m^{\dag} ( \partial_\tau  + \tilde{\epsilon}_f  ) z_m \nonumber \\
& &+ \frac{1}{N}\,\int_0^\beta d \tau d \tau^\prime \sum_m \Sigma_0(\tau-\tau^\prime)\,
 z_m^{\dag}(\tau) R(\tau)  z_m(\tau^\prime) R(\tau^\prime) \nonumber \\
& &+\int_0^\beta d\tau \frac{M}{2}\Big((\partial_\tau Q)^2+  \Omega^2 Q^2 \Big)  \nonumber \\
& &+\int_0^\beta d\tau \,g_0 Q \,\Big(  \sum_m z^\dag_m z_m
- \langle \sum_m z^\dag_m z_m \rangle \Big)\nonumber \\
& &+\int_0^\beta d\tau \,i (\frac{d \theta}{d \tau}+\Omega_0)(\sum_m z_m^\dag z_m + R^2-q N)
\nonumber \\
& &+\int_0^\beta d\tau \,\lambda_{sa}(R^2-q N),
\ea
where
\be
\delta \tilde{\epsilon}_f=\epsilon_f+\lambda_{sa},\;\; q=1/N.
\ee
The integration of phonon $Q$  contributes the following term to the action
\ba
\label{ac:phononcont}
& &-\frac{g_0^2}{2 M}\,\int_{0}^\beta d \tau\,\int_{0}^\beta d \tau^\prime\,
T \sum_{i\omega}\, \frac{e^{-i \omega (\tau-\tau^\prime)}}{\omega^2+\Omega^2}\nonumber \\
& &\times \sum_m \delta  n_{sm}(\tau)  \sum_n \delta n_{sn}(\tau^\prime),
\ea
where
$ \delta n_{sm}=s_m^\dag s_m-\langle s_m^\dag s_m \rangle$.
In order for the $1/N$ expansion to be applicable,
each term of the action should be proportional to $N$.
The structure of Eq.~(\ref{ac:phononcont}) suggests that the electron-phonon coupling $g_0$ should be
{\it rescaled} as follows:
\be
\label{scale1}
g_0=g/\sqrt{N}.
\ee
By the same reason $R(\tau)$ should be also rescaled.
\be
\label{scale2}
R(\tau)= \sqrt{N} \, r(\tau).
\ee
The rescaled action becomes
\ba
\label{action:rn3}
& &S=\int_0^\beta d \tau \,
\sum_m z_m^{\dag} ( \partial_\tau  + \tilde{\epsilon}_f  ) z_m \nonumber \\
& &+\int_0^\beta d \tau d \tau^\prime \sum_m \Sigma_0(\tau-\tau^\prime)\,
 z_m^{\dag}(\tau) r(\tau)  z_m(\tau^\prime) r(\tau^\prime) \nonumber \\
& &+\int_0^\beta d\tau \frac{M}{2}\Big((\partial_\tau Q)^2+  \Omega^2 Q^2 \Big)  \nonumber \\
& &+\frac{1}{\sqrt{N}}\,\int_0^\beta d\tau \,g Q \,\Big(  \sum_m z^\dag_m z_m
- \langle \sum_m z^\dag_m z_m \rangle \Big)\nonumber \\
& &+\int_0^\beta d\tau \,i (\frac{d \theta}{d \tau}+\Omega_0)\Big(\sum_m z_m^\dag z_m   \Big)
\nonumber \\
& &+N \int_0^\beta d\tau \,i (\frac{d \theta}{d \tau}+\Omega_0)(r^2-q) \nonumber \\
& &+N\int_0^\beta d\tau \,\lambda_{sa}(r^2-q ),
\ea
Now the $1/N$ expansion can be directly applied to the action Eq.~(\ref{action:rn3}).
%%%%%%%%%%%%%%%%%%%%%%%%%%%%%%%%%%%%%%%%%%%%%%%%%%%%%%%%%%%%%%%%%%%%%%%%%%%%
\section{Saddle point approximation}
The leading order approximation of the $1/N$ expansion amounts to
the saddle point approximation or equivalently, 
the mean-field approximation. The mean-field theory of infinite-$U$ Anderson
model has been studied in great detail in Sec.\ II of Coleman's
paper.\cite{coleman} Here, we briefly summarize the main results of
Ref.\cite{coleman} for the purpose of the self-contained treatment
and the introduction of necessary notations.

We assume that $r, \lambda$ take a \textit{static} value $r_{sa}, \lambda_{sa}$
in the saddle point approximation, respectively.
In the saddle point approximation the following flucutations with respect to the saddle point value
\be
\label{scale3}
i (\frac{d \theta}{d \tau}+\Omega_0),\;\; \delta r = r - r_{sa}
 \ee
are neglected. We will treat the electron-phonon interaction on an equal footing with the above fluctuations, so that
the electron-phonon interaction is \textit{not} included in the saddle point approximation.
Now the action appropriate for the saddle point approximation can be expressed as
\ba
\label{action:rn4}
& &S_{mf}=\int_0^\beta d \tau \,
\sum_m z_m^{\dag} ( \partial_\tau  + \tilde{\epsilon}_f  ) z_m \nonumber \\
& &+\int_0^\beta d \tau d \tau^\prime \sum_m \Sigma_0(\tau-\tau^\prime)\,r_{sa}^2
 z_m^{\dag}(\tau)  z_m(\tau^\prime)  \nonumber \\
& &+N\, \int_0^\beta d\tau \,\lambda_{sa}(r_{sa}^2-q ),
\ea
Next we integrate over  $z_m$ to obtain the saddle point action $\tilde{S}_{sa}(r_{sa},\lambda_{sa})$ which is
\be
e^{- \tilde{S}_{sa}(r_{sa},\lambda_{sa})}=\int D[z_m]\,e^{-S_{mf}(z,r_{sa},\lambda_{sa})}.
\ee
The explicit form of $\tilde{S}_{sa}$ is given by
\ba
\tilde{S}_{sa}&=&N\int_0^\beta d\tau \,\lambda_{sa}(r_{sa}^2-q ) \nonumber \\
&-&N \textrm{Tr}
\ln [ \partial_\tau + \tilde{\epsilon}_f + r^2_{sa} \Sigma_0 ].
\ea
The  saddle-point values $r_{sa}, \lambda_{sa}$ are
determined by minimizing $\tilde{S}_{sa}$  with respect to them.
The minimization with respect to $\lambda_{sa}$ yields
\be
\label{eq:spa1}
r_{sa}^2-q=- T \sum_{i\epsilon} \frac{1}{i\epsilon-\epsilon_f-\lambda_{sa}+ i \Delta {\rm sgn}(\epsilon)},
\ee
where
\be
\Delta  \equiv \Delta_0 r_{sa}^2.
\ee
At zero temperature the summation of Eq.~(\ref{eq:spa1}) can be done in a closed form.
\be
\label{eq:spa3}
r_{sa}^2-q = - \Big[ \half - \frac{1}{\pi} \tan^{-1} \big( \frac{\epsilon_f +\lambda_{sa}}{\Delta} \big) \Big ].
\ee
Likewise, the minimization with respect to $r_{sa}^2$ yields
\be
\label{eq:spa2}
\lambda_{sa}=-T \sum_{i\epsilon}\,
\frac{- i\Delta_0 {\rm sgn}(\epsilon)}{i\epsilon-\epsilon_f-\lambda_{sa}+ i \Delta {\rm sgn}(\epsilon)}.
\ee
At zero temperature the summation of Eq.~(\ref{eq:spa2}) can be done in a closed form.
\be
\label{eq:spa4}
\lambda_{sa}=\frac{\Delta_0}{2\pi}\,\ln \left[ \frac{D^2}{\Delta^2+(\epsilon_f+\lambda_{sa})^2} \right],
\ee
where $D$ is an energy cutoff which is the order of the bandwidth of conduction electrons.
Two real-valued equations Eq.~(\ref{eq:spa3}) and Eq.~(\ref{eq:spa4}) can be combined into a single
complex-valued equation.
\be
\label{eq:sad}
\frac{\Delta_0}{\pi}\,\ln \left[ \frac{\pi \xi}{\Delta_0} \right ]+ \xi = \epsilon_f^*+ i \Delta_0 q,
\ee
where
\ba
& &\xi=\epsilon_f + \lambda_{sa} + i \Delta,\nonumber \\
& &\epsilon_f^*= \epsilon_f + \frac{\Delta_0}{\pi} \ln \left[ \frac{\pi D}{\Delta_0} \right ].
\ea
The Kondo limit is specified by the condition $|\xi | \ll \Delta_0$, and in this limit the first term of
the right hand side of Eq.~(\ref{eq:sad}) dominates.
In the Kondo limit,  the Kondo temperature scale  is given by
\be
\label{def:Kondo}
\Delta_K \equiv |\xi|=\sqrt{\Delta^2 + \tilde{\epsilon}_f^2} \sim D e^{ \pi \epsilon_f/\Delta_0}.
\ee
This result derived from the saddle point approximation
 coincides with the one obtaind from the scaling theory of Anderson
model by Haldane except for the prefactor.\cite{haldane,note1}
A few numerical solutions of Eqs.~(\ref{eq:spa3},\ref{eq:spa4})are presented in Table I.
\begin{table}
\label{table:spa}
\begin{tabular}[c]{|c|c|c|c|}
 \hline
  $q$  & $\Delta/\Delta_0 $    & $\tilde{\epsilon}_f/\Delta_0$ & $\Delta_K/\Delta_0$ \\ \hline
  1/2 & $2.80 \times 10^{-3} $ & $ 2.46 \times 10^{-5}$  & $2.80 \times 10^{-3}$ \\ \hline
  1/4 & $1.95 \times 10^{-3} $ & $ 1.99 \times 10^{-3}$ & $2.79 \times 10^{-3}$ \\ \hline
  1/6 & $1.39 \times 10^{-3} $ & $2.41 \times 10^{-3} $ & $2.78 \times 10^{-3}$ \\ \hline
\end{tabular}
\caption{The numerical solutions of the saddle point equations
Eqs.~(\ref{eq:spa3},\ref{eq:spa4}). The input
parameters are $D= 65 \Delta_0$ and $\epsilon_f= - 3.2 \Delta_0$.
$\Delta_K$ is defined in Eq.~(\ref{def:Kondo}). }
\end{table}
From the  numerical solutions we find
that in the case of $q=1/2$,
$\Delta \gg  \tilde{\epsilon}_f$, while for  $q=1/4$ and $1/6$,  $\Delta \sim \tilde{\epsilon}_f$.
In all   cases $\Delta_K $ and $\Delta$ are of the same order of magnitude.
%%%%%%%%%%%%%%%%%%%%%%%%%%%%%%%%%%%%%%%%%%%%%%%%%%%%%%%%%%%%%%%%%%%%%%%%%%%%%
\section{$1/N$ fluctuations}
The $1/N$ corrections are related to the fluctuations with respect to the saddle point configuration.
It is convenient to treat the fluctuations collectively, so let us define a two-component vector $X$ as
follows:
\ba
X&=&\begin{pmatrix} \delta r \cr i r_{sa} \Theta \end{pmatrix}, \nonumber \\
\Theta &=& \frac{d \theta}{d \tau}+\Omega_0.
\ea
 The $1/N$ corrections can be computed systematically
by expanding the action Eq.~(\ref{action:rn3}) with respect to the fluctuations and by
integrating out $z_m$.
The action Eq.~(\ref{action:rn3}) can be re-expressed as follows.
\ba
\label{action:rn2}
S&=&S_{X}^{(0)}+S_z+S_{ph}+\delta S_z, \\
S_{X}^{(0)}&=& \frac{N}{2} \int_0^\beta d \tau\, X^T \Gamma^{(0)} X, \\
S_z&=& \sum_{m,i\epsilon}\,\Big[-i\epsilon+\tilde{\epsilon}_f
- i \Delta {\rm sgn}(\epsilon) \Big]\,
z_m^\dag z_m, \\
S_{ph} &=& \int_0^\beta d\tau \frac{M}{2}\Big((\partial_\tau Q)^2+  \Omega^2 Q^2 \Big), \\
\delta S_z &=& \sum_m \Big[ \int_0^\beta d\tau\,\Big( i \Theta + \frac{g Q}{\sqrt{N}}+\phi_m \Big) z_m^\dag z_m  \nonumber \\
&-& \int_0^\beta d\tau\, \frac{g Q}{\sqrt{N}} \langle z_m^\dag z_m  \rangle \Big]  \nonumber \\
&+& \sum_m  \int_0^\beta d\tau \int_0^\beta d\tau^\prime\, \Sigma_0(\tau-\tau^\prime) z_m^\dag(\tau)
z_m(\tau^\prime) \nonumber \\
&\times& \Big( r_{sa} \delta r (\tau^\prime) + r_{sa} \delta r (\tau) + \delta r (\tau) \delta r (\tau^\prime) \Big),
\ea
where
\be
 \Gamma^{(0)} = 2 \begin{pmatrix}
  \lambda_{sa}  & 1  \cr   1  & 0 \end{pmatrix}.
\ee
$\phi_m$ is the source field necessary for the computation of impurity susceptibilities.
The irrelevant constants are omitted  in Eq.~(\ref{action:rn2}).

Next step is to integrate over $z_m$. This step involves $S_z$ and $\delta S_z$ of the action Eq.~(\ref{action:rn2}).
To simplify notations, define
\be
G^{-1}_0(i\epsilon)=i\epsilon + i \Delta \, {\rm sgn}(\epsilon) - \tilde{\epsilon}_f.
\ee
The integration over $z_m$ generates
\be
\label{logdet}
-\sum_m {\rm Tr} \ln \Big[-G^{(0)}(\tau-\tau^\prime)+ M(\tau,\tau^\prime) \Big ],
\ee
where $M(\tau,\tau^\prime)$ is given by
\ba
& &M(\tau,\tau^\prime)=\big( \phi_m + \frac{g}{\sqrt{N}} Q + i \Theta \big )(\tau) \delta (\tau-\tau^\prime)  \nonumber \\
& & +\Sigma_0(\tau-\tau^\prime)
\big( r_{sa} \delta r(\tau) + r_{sa} \delta r(\tau^\prime) +\delta r(\tau)  \delta r(\tau^\prime) \big ).
\ea
In our approximation we expand  the logarithm of Eq.~(\ref{logdet}) with respect to $M(\tau,\tau^\prime)$
up to the second order. This is equivalent to reckoning in the Gaussian fluctuations but not the
higher order fluctuations in $1/\sqrt{N}$ (see Appendix \ref{1/N}).

The relevant part of the expansion is
$$+ \sum_m {\rm Tr} \Big[ G_0 M \Big] + \half  \sum_m \Big[ G_0 M G_0 M \Big].$$
The details of further calculations are presented in the appendix \ref{append:1/N}.
The effective action of $X, Q, \phi_m$ is given by
\ba
\label{action:general}
S_{eff}^{(a)}&=& \frac{N}{2}\,  \sum_{i\omega} \, X^T (-i\omega)\,\Gamma^{(1)}\, X(i\omega) \nonumber \\
&+&\sum_m \sum_{i \omega}\, \frac{g}{\sqrt{N}}\,(r_{sa} Q)(i\omega) \nonumber \\
&\times &\Big( K_{\theta \theta}^{(1)}( i \theta r_{sa} + \phi_m r_{sa}) + K_{r \theta}^{(1)}\, \delta r\Big)(-i\omega) \nonumber \\
&+&\sum_m \sum_{i \omega}\, K_{\theta \theta}^{(1)}(i\omega)\,( i \theta r_{sa})(i\omega)\,(\phi_m r_{sa})(-i \omega) \nonumber \\
&+&\half \sum_m \sum_{i \omega}  K_{\theta \theta}^{(1)}(i\omega)\, (\phi_m r_{sa})^2 \nonumber \\
&+& \sum_m \sum_{i \omega} K_{r \theta}^{(1)}\,\delta r (i \omega)\, r_{sa} \phi_m(-i\omega) \nonumber \\
&+&\half \sum_{i \omega}(M\omega^2+M \Omega^2+ K_{\theta \theta}^{(1)}\,g^2 r^2_{sa}) Q(i \omega) Q(-i \omega),
\ea
where
\be
\Gamma^{(1)}(i\omega)= \begin{pmatrix}
2 \lambda_{sa}+ \sum_{i=0,2,3} K_{rr}^{(i)}  &  2 + K_{r \theta}^{(1)} \cr
 2 + K_{r \theta}^{(1)} & K_{\theta \theta}^{(1)}
 \end{pmatrix}.
\ee
The explicit forms of various polarization functions $K$ can be found in Appendix \ref{polarization}.
By the saddle point condition Eq.~(\ref{eq:spa2}),
\be
2 \lambda_{sa}+K_{rr}^{(0)}=0.
\ee
Then it follows that
\be
\label{gamma1}
\Gamma^{(1)}(i\omega)= \begin{pmatrix}
 K_{rr}^{(2)}+K_{rr}^{(3)} &  2 + K_{r \theta}^{(1)} \cr
 2 + K_{r \theta}^{(1)} & K_{\theta \theta}^{(1)}
 \end{pmatrix}.
\ee
In fact, only $\textrm{Re} \, K_{rr}^{(2)}$ in the (1,1) element of Eq.~(\ref{gamma1})
 contributes to the action due to the symmetry $\omega \leftrightarrow -\omega$.
Now the integration  over phonon
\be
\exp[-S_{eff}^{(b)}]=\int D[Q]\,\exp[-S_{eff}^{(a)}(X,Q,\phi_m)]
\ee
contributes the following to $S_{eff}^{(b)}$.
\ba
& &-\frac{E_p}{2 N}\, r_{sa}^2\, \frac{\Omega^2}{\omega^2+ \Omega^2+K_{\theta \theta}^{(1)} g^2 r_{sa}^2}
\nonumber \\
& &\times \left| \sum_m \Big( K_{\theta \theta}^{(1)} r_{sa}( i \Theta + \phi_m)(i \omega) +
K_{r \theta}^{(1)} \delta r(i \omega)  \Big) \right |^2,
\ea
where
\be
E_p = \frac{ g^2}{ M \Omega^2}
\ee
is the polaron energy. Since we are assuming that the  electron-phonon correlation is weak, the relation
\be
E_p < \Delta_0
\ee
holds.
We will also assume that $E_p > \Delta_K$, which is more relavant to the real physical systems.
The effect of electron-phonon interaction can be traced by following  $E_p$.
Let us define the following dimensionless function
\be
\label{dfunction}
D(i \omega) = \frac{\Omega^2}{\omega^2+ \Omega^2+K_{\theta \theta}^{(1)} g^2 r_{sa}^2/M}.
\ee
The explicit form of  the effective action $S_{eff}^{(b)}(X,\phi_m)$ is given by
\ba
S_{eff}^{(b)}&=&\frac{N}{2}\,\sum_{i \omega} X^T(-i\omega)  \Gamma(i\omega)\, X(i\omega) \nonumber \\
&+& \half \sum_m \sum_{i \omega} K_{\theta \theta}^{(1)}(i \omega)
r_{sa}^2 \phi_m(i\omega) \phi_m(-i\omega) \nonumber \\
&-& \frac{E_p r_{sa}^4}{2 N} \sum_{m,n,i\omega}\,\big[ K_{\theta \theta}^{(1)} \big]^2 D(i\omega)\,
 \phi_m (i\omega) \phi_n(-i \omega)  \nonumber \\
&+&\sum_m \sum_{i \omega}\,\big(1-E_p D(i\omega) r_{sa}^2 K_{\theta \theta}^{(1)} \big) \nonumber \\
&\times& r_{sa} \phi_m(i\omega)\,
\Big( K_{\theta \theta}^{(1)} r_{sa} i \Theta(-i \omega) + K_{r \theta}^{(1)}\,\delta r(-i \omega) \Big),
\ea
where
\ba
\Gamma(i\omega)&=&\begin{pmatrix}    K_{rr} & K_{r \theta} \cr K_{r \theta} &  K_{\theta \theta}
\end{pmatrix}, \nonumber \\
K_{rr}&=&K_{rr}^{(1)}-E_p r_{sa}^2 D (i \omega) [ K_{r \theta}^{(1)} ]^2, \nonumber \\
K_{r\theta}&=&2+K_{r \theta}^{(1)}-E_p r_{sa}^2  D (i \omega)\,
K_{r \theta}^{(1)} K_{\theta \theta}^{(1)},  \nonumber \\
K_{\theta \theta}&=&K_{\theta \theta}^{(1)} \Big(1-E_p D(i\omega) r_{sa}^2 K_{\theta \theta}^{(1)} \Big).
\ea
Finally the integration over $X$
\be
\exp[-S_{eff}^{(c)}(\phi_m)]=\int D[X]\,\exp[-S_{eff}^{(b)}(X,\phi_m)]
\ee
yields the effective action of $\phi_m$.
\ba
\label{finalaction}
S_{eff}^{(c)}&=& \half \sum_m \sum_{i \omega} K_{\theta \theta}^{(1)}(i \omega)
r_{sa}^2 \phi_m(i\omega) \phi_m(-i\omega) \nonumber \\
&-& \frac{E_p r_{sa}^4}{2 N} \sum_{m,n,i\omega}\,\big[ K_{\theta \theta}^{(1)} \big]^2 D(i\omega)\,
 \phi_m (i\omega) \phi_n(-i \omega)  \nonumber \\
&-& \frac{ r_{sa}^2}{2N}\,\sum_{m,n,\omega}\,
\big(1-E_p D(i \omega) r_{sa}^2 K_{\theta \theta}^{(1)} \big )^2  \nonumber \\
&\times &\phi_m(i\omega) \phi_n(-i \omega) [K_{r \theta}^{(1)}, K_{\theta \theta}^{(1)}] \Gamma^{-1}\,
\begin{bmatrix} K_{r \theta}^{(1)} \cr K_{\theta \theta}^{(1)} \end{bmatrix}.
\ea

%%%%%%%%%%%%%%%%%%%%%%%%%%%%%%%%%%%%%%%%%%%%%%%%%%%%%%%%%%%%%%%%%%%%
\section{The spin and the charge  susceptibilities of impurity}
\label{sec:susc}
The spin   susceptibility of impurity is defined by
\ba
\chi_s(\tau-\tau^\prime)&=& \langle M_z(\tau) \, M_z(\tau^\prime)  \rangle, \\
M_z &=& \sum_m \, m s_m^\dag s_m=\sum_m\,m z_m^\dag z_m,
\ea
where the sum over $m$ runs from $-N/2$ to $N/2$.
The charge susceptibility is  defined by
\be
\chi_c(\tau-\tau^\prime)= \langle \delta \hat{n}_f(\tau) \,\delta \hat{n}_f(\tau^\prime)\rangle,
\ee
where
\ba
\delta \hat{n}_f &=&  \sum_m  s_m^\dag s_m - \langle \sum_m  s_m^\dag s_m \rangle \nonumber \\
&=& \sum_m  z_m^\dag z_m - \langle \sum_m  z_m^\dag z_m \rangle
\ea
The spin and charge susceptibilites can be computed from
\be
\label{formula1}
\chi_{m n}(\tau,\tau^\prime)=
\frac{\delta^2 \ln Z}{\delta \phi_m(\tau)\, \delta \phi_{n}(\tau^\prime)} \Big|_{\phi \to 0}.
\ee
Here the partition function $Z$ is understood to be a functional of $\phi_m$ only, namely
all other fields should be integrated out.
In our approximation,
\be
Z=\int D[\phi_m]\, \exp[-S_{eff}^{(c)}(\phi_m)],
\ee
where $S_{eff}^{(c)}(\phi_m)$ is given by Eq.~(\ref{finalaction}).
$\chi_{m n}(\tau,\tau^\prime)$ in frequency space is given by
\be
\chi_{m n}(\tau,\tau^\prime)= T \sum_{i \omega}\, \chi_{mn}(i\omega)\,e^{-i \omega (\tau-\tau^\prime)}.
\ee
Then it follows that
\ba
\chi_c(i \omega)&=&\sum_{m,n}\,\chi_{m n}(i\omega),\nonumber \\
\chi_s(i \omega )&=&\sum_{m,n}\,m n \, \chi_{m n}(i \omega).
\ea
From  Eq.~(\ref{formula1}) we find
\ba
\label{chimn}
\chi_{mn}&=&-\delta_{mn}\,r_{sa}^2\, K_{\theta \theta}^{(1)}( i\omega) \nonumber \\
&+&\frac{r_{sa}^2}{N}\, \big(1-E_p D(i \omega) r_{sa}^2 K_{\theta \theta}^{(1)} \big )^2 \nonumber \\
&\times& \big[K_{r \theta}^{(1)}, K_{\theta \theta}^{(1)}\big] \Gamma^{-1}
\begin{bmatrix} K_{r \theta}^{(1)} \cr K_{\theta \theta}^{(1)} \end{bmatrix} \nonumber \\
&+&\frac{r_{sa}^4}{N}\,E_p\,D(i\omega)\,\Big[ K_{\theta \theta}^{(1)} ( i\omega) \Big ]^2.
\ea
From Eq.~(\ref{chimn}) the spin susceptibility is readily obtained
\ba
\label{spinsusc1}
\chi_s(i \omega ) &=& \big[ \sum_{m} m^2 \big ]\, \tilde{\chi}_s(i\omega), \nonumber \\
\tilde{\chi}_s(i\omega) &=& -r_{sa}^2\, K_{\theta \theta}^{(1)}( i\omega).
\ea
In our approximation scheme the spin susceptibility is {\it not} renormalized by the electron-phonon
interaction.
Clearly this is an artifact of the approximation of the saddle point and the Gaussian fluctuation.
If the fluctuations beyond the Gaussian fluctuations ( higher order in $1/\sqrt{N}$) are considered
the spin susceptibility is expected to be renormalized by the electron-phonon. However, the corrections due to
higher order fluctuations will be rather small.
It is illuminating to compare this result with that of numerical renormalization group studies.
Hewson and Meyer have computed the imaginary part of the retarded spin susceptibility
for the \textit{symmetric and finite}-$U$ AH model
(see the Fig. 13 of Ref. \onlinecite{hewson}. Compare two cases with $\lambda=0.0$ and $\lambda=0.02$).
Their result shows that the electron-phonon interaction
does not influence the qualitative behavior of the spin susceptibility except for the decrease of
the peak height at very low  energy around  the Kondo temperature scale.  
For our parameter regime where the electron correlation is far more dominant
over the electron-phonon interaction, the decrease of the peak height would be hardly noticeable.

The charge susceptibility $\chi_c(i \omega )$ is given by
\ba
\label{chargeoriginal}
& &\chi_c(i \omega )/N = r_{sa}^2 \Big(1-E_p D(i \omega) r_{sa}^2 K_{\theta \theta}^{(1)} \Big )^2 \nonumber \\
& &\times \big[K_{r \theta}^{(1)}, K_{\theta \theta}^{(1)}\big] \Gamma^{-1}
\begin{bmatrix} K_{r \theta}^{(1)} \cr K_{\theta \theta}^{(1)} \end{bmatrix} \nonumber \\
& &-r_{sa}^2\, K_{\theta \theta}^{(1)}( i\omega)
\Big(1-E_p D(i \omega) r_{sa}^2 K_{\theta \theta}^{(1)} \Big ) \nonumber \\
& & \equiv \tilde{\chi}_c(i\omega).
\ea
The charge susceptibility is indeed renormalized by the electron-phonon interaction in a very complicated way.
This complicated expression can be recast in a more transparent form in the following way.
First note that the charge susceptibility   \textit{in the absence of} the electron-phonon interaction
is given by
\be
\label{susc.cha.abs}
\tilde{\chi}_c^{(0)}= r_{sa}^2 \Big(
\big[K_{r \theta}^{(1)}, K_{\theta \theta}^{(1)}\big] \big[\Gamma^{(1)} \big]^{-1}
\begin{bmatrix} K_{r \theta}^{(1)} \cr K_{\theta \theta}^{(1)} \end{bmatrix}
 -K_{\theta \theta}^{(1)} \Big ).
\ee
This result can be obtaind from Eq.~(\ref{chargeoriginal}) by putting $E_p=0$ everywhere and was derived by
Coleman.\cite{coleman}
Explicit calculation shows that
\be
\tilde{\chi}_c^{(0)}(i\omega)=4 r_{sa}^2
\left[ \frac{ K_{\theta \theta}^{(1)} }{ K_{r r}^{(1)}K_{\theta \theta}^{(1)}-
(K_{r \theta}^{(1)}+2)^2} \right ].
\ee
For the subsequent analyses it is helpful to examine the susceptibilities in the absence of phonons in more detail.
The susceptibilities at zero frequency in the absence of phonons are given by
\ba
\label{susczero}
\tilde{\chi}^{(0)}_s(i\omega = 0) &=& \frac{1}{\pi}\,\frac{\Delta}{\Delta_K^2}, \nonumber \\
\tilde{\chi}^{(0)}_c(i\omega = 0) &=& \frac{\pi \Delta}{(\Delta_0 + \pi \tilde{\epsilon}_f)^2+(\pi \Delta)^2}.
\ea
In the Kondo regime, the relation $\Delta_0 \gg  \pi \sqrt{\Delta^2 + \tilde{\epsilon}^2_f}$ holds,
so that we have
\be
\tilde{\chi}^{(0)}_c(\omega = 0) \sim \frac{\pi \Delta}{ \Delta_0^2}.
\ee
The ratio of the charge and spin susceptibility at zero frequency is then given by
\be
\label{theratio}
\tilde{\chi}^{(0)}_c(i\omega = 0)/\tilde{\chi}^{(0)}_s(i\omega = 0)
\sim \left(\frac{\pi \Delta_K}{\Delta_0} \right )^2 \ll 1.
\ee
In fact, over  most of the frequency range of interest, the absolute magnitude of spin susceptibility is far
greater than that of the charge susceptibility. This is because the charge fluctuations are severely suppressed by the
very strong local Coulomb repulsion in the low temperature Kondo regime,
while the spin fluctuations (spin flips) are the dominant processes there.

%%%%%%%%%%%%%%%%%%
\begin{figure}
\scalebox{0.7}{\includegraphics{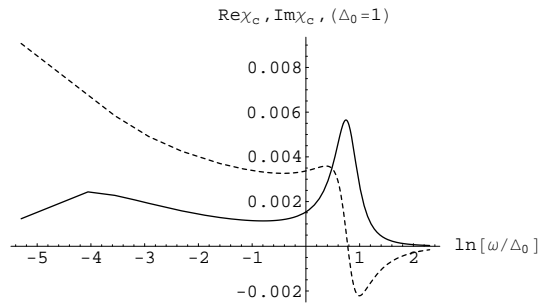}}
\caption{\label{fig:susc} The real and imaginary part of the retarded
charge susceptibility in real frequency in the absence of electron-phonon
interaction $\tilde{\chi}^{R,(0)}_c(i\omega \to \omega+i \delta)$.
The dashed line and the solid line indicates  the real part and the  imaginary part, respectively.
Note that the frequency is in the logarithmic scale, and the y-axis is in the absolute scale.
The input parameters are identical with those of Table I. The value of $q$ is 1/2.}
\end{figure}
%%%%%%%%%%%%%%%%%%

We also note that $\tilde{\chi}_c^{(0)}(i\omega)$ is related to the propagator
$\langle \delta r(i\omega) \delta r(-i \omega) \rangle^{(0)} $ in the absence of
phonon in the following way
\be
\langle \delta r(i\omega) \delta r(-i \omega) \rangle^{(0)}=\frac{1}{N}\,
\frac{1}{ 4 r_{sa}^2 }\,\tilde{\chi}_c^{(0)}(i\omega),
\ee
which can be shown by inverting the matrix $\Gamma^{(1)}(i\omega)$.
This implies that  higher order corrections stemming from
 the bose fluctuations are suppresed in the limit of large $N$.

Using the above results on the susceptibilities in the absence of phonon and
expanding the matrix $\Gamma^{-1}$ up to the second order of $g$, the result Eq.~(\ref{chargeoriginal})
\textit{in the presence of phonon} can be
expressed in a more transparent form:
\be
\label{charge1}
\tilde{\chi}_c(i\omega) \approx \tilde{\chi}^{(0)}_c(i\omega)\,\Big[
1+E_p D(i\omega)\,\tilde{\chi}^{(0)}_c(i\omega) \Big],
\ee
where $D(i\omega)$ is defined by Eq.~(\ref{dfunction}).
Using the previous results we find
\be
\left| E_p D(i\omega)\,\tilde{\chi}^{(0)}_c(i\omega) \right | \ll 1.
\ee
The structure of Eq.~(\ref{charge1}) clearly prompts us to write
\be
\label{charge2}
\tilde{\chi}_c(i\omega)
 \sim  \frac{\tilde{\chi}^{(0)}_c(i\omega)}{1-E_p D(i\omega)\,\tilde{\chi}^{(0)}_c(i\omega)}.
\ee
The imaginary part of the retarded charge susceptibility [Eq.~(\ref{charge1})] in real frequency can be expressed as
\ba
\textrm{Im} \tilde{\chi}_c^{R}(\omega) &\sim& \textrm{Im} \tilde{\chi}^{(0),R}_c(\omega) \nonumber \\
&\times&\Big(
1+E_p\,\textrm{Re}[D^R(\omega)]\,\textrm{Re}[ \tilde{\chi}_c^{R,(0)}(\omega)] \Big),
\ea
where only the leading contribution is exhibited.
As depicted in Fig. \ref{fig:susc},  the variations of $\textrm{Re}[ \tilde{\chi}_c^{R,(0)}(\omega)]$ is almost negligible
for all frequency region below the frequency of the order of $|\epsilon_f|$. Therefore, the major frequency dependence of
the electron-phonon correction to the charge susceptibility originates from $\textrm{Re}[D^R(\omega)]$:
\be
\label{theD}
\textrm{Re}[D^R(\omega)] \sim
\frac{\Omega^2}{-\omega^2+\Omega^2},
\ee
where the singularity at $\omega = \Omega$ is cut-off by the imaginary part of $K^{(1)}_{\theta \theta}$ but this is
irrelevant in our discussion.
In view of Eq.~(\ref{theD}) a qualitative change near the phonon frequency $\Omega$ is expected.
More specificially,  one can expect
 that the imaginary part of the charge susceptibility
is to be enhanced at frequency slightly lower than the phonon frequency while it is expected to be reduced at frequency
slightly higher than the phonon frequency. This expectation is in good agreement with the results of NRG studies by
Hewson and Meyer
(see the left panel of Fig. 13 in Ref. \onlinecite{hewson} and compare two cases of $\lambda=0.0$ and $\lambda=0.02$).

%%%%%%%%%%%%%%%%%%
\begin{figure}
\scalebox{0.7}{\includegraphics{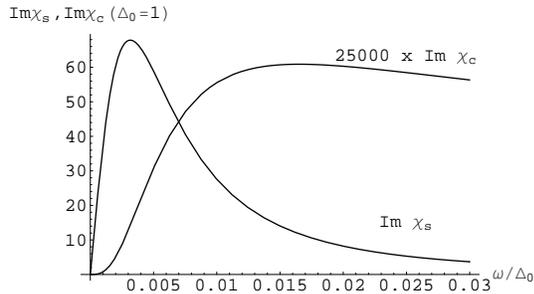}}
\caption{\label{fig:twosusc} The  imaginary parts of the
retarded charge susceptibility and spin susceptibility in the absence of electron-phonon
interaction. The charge susceptibility has been multiplied by a factor of 25,000 to make comparison
easier. The input parameters are identical with those of Table I. $q=1/2$.}
\end{figure}
%%%%%%%%%%%%%%%%%%
%%%%%%%%%%%%%%%%%%
\begin{figure}
\scalebox{0.7}{\includegraphics{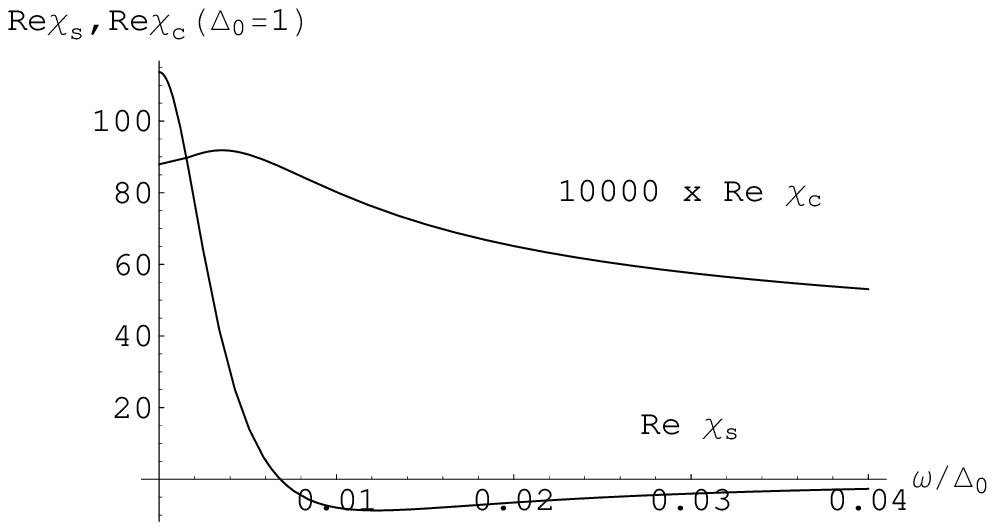}}
\caption{\label{fig:twosuscre} The  real parts of the
retarded charge susceptibility and spin susceptibility in the absence of electron-phonon
interaction. The charge susceptibility has been multiplied by a factor of 10,000 to make comparison
easier. The input parameters are identical with those of Table I. $q=1/2$.}
\end{figure}
%%%%%%%%%%%%%%%%%%

The charge susceptibility at zero frequency in the presence of phonon  can be expressed
\be
\label{main1}
\tilde{\chi}_c(i\omega = 0) \sim \frac{\pi \Delta}{\Delta_0^2}\,
\Big(1+ \frac{\pi E_p \Delta}{\Delta_0^2} \Big).
\ee
For the case of $q=1/2$  the relation $\Delta_K \sim \Delta$ holds, and
the result Eq.~(\ref{main1}) can be interpreted from the viewpoint of
the renormalization of Kondo temperature by electron-phonon interaction.
\be
\label{main2}
T_K \sim T_K^{(0)}\,\Big(1+ \frac{\pi E_p \Delta}{\Delta_0^2} \Big),
\ee
where $T_K^{(0)} = \Delta_K$ is the Kondo temperature in the absence of phonons.
The Eq.~(\ref{main2}) indicates that the electron-phonon interaction {\it increases} the
Kondo temperature, which is consistent with the recent results found by NRG method.\cite{hewson,jpc}.
Since $E_p < \Delta_0$ and $\Delta \ll \Delta_0$ the amount of increase is rather small.
This smallness of renormalization is again due to the suppressed charge fluctuations.
With Coulomb repulsion fixed to be infinite from the outset, the charge fluctuation is controlled by the
$\epsilon_f$. As the impurity level goes down deeper into the Fermi sea of conduction electron,
evidently the charge fluctuation is reduced. As can be seen in Eq.~(\ref{def:Kondo}), $\Delta$ decreases \textit{exponentially}
with the decreasing $\epsilon_f$ (in the sense of becoming more negative).

%%%%%%%%%%%%%%%%%%%%%%%%%%%%%%%%%%%%%%%%%%%%%%%%%%%%%%%%%%%%%%%%
\section{Phonon Propagator}
The phonon propagator can be obtained from the action Eq.~(\ref{action:general}) by setting
the source field $\phi_m$ to zero and then by integrating over the bose fluctuations $X$ and the fermions $z_m$.
\be
\label{phononprop1}
D_{ph}(i \omega)=\Big[ M \omega^2+ M \Omega^2 - g^2 \tilde{\chi}_c^{(0)}(i \omega) \Big]^{-1}.
\ee
The physics of Kondo fixed point is built in $\tilde{\chi}_c^{(0)}(i \omega)$.
If the  bose fluctuations $X$ were \textit{not} taken into account, the phonon propagator would be
\be
\label{wouldbe}
 \Big[ M \omega^2+ M \Omega^2 - g^2 \tilde{\chi}_s^{(0)}(i \omega) \Big]^{-1}.
\ee
This can be seen from Eq.~(\ref{action:general}) by setting the bose fluctuation $X$
and the  source field $\phi_m$ to zero, and then by integrating over $z_m$.
As discussed in Sec.~\ref{sec:susc} and can be seen Fig. \ref{fig:twosusc} and Fig. \ref{fig:twosuscre}
the charge and spin susceptibility of impurity behave essentially
differently from each other. 
First of all, the absolute magnitude of the spin susceptibility is much larger than
that of the charge susceptibility by the order of magnitudes. 
Second the frequency dependences are also different from each other.
The imaginary part of the spin susceptibility 
$\textrm{Im} \tilde{\chi}_s^{R,(0)}(\omega)$ has a sharp peak at $\omega \sim \Delta_K$, while the
imaginary part of the charge susceptibility
$\textrm{Im} \tilde{\chi}_c^{R,(0)}(\omega)$ has a rather broad peak at frequency higher than $\Delta_K$.
$\textrm{Im} \tilde{\chi}_c^{R,(0)}(\omega)$ also has a peak at high frequency 
order of $-\epsilon_f$ as seen in Fig.~\ref{fig:susc}.

 If the bose fluctuations were not considered,  then
based on Eq.~(\ref{wouldbe}) one would anticipate rather substantial renormalization even for the relatively small values
of $g$. This is in stark contrast with the results by NRG\cite{hewson,jpc} which show rather small renormalization.
To make arguments more concrete
let us compare last two terms in the bracket of Eq.~(\ref{wouldbe}) at zero frequency.
Using the result Eq.~(\ref{susczero}) we obtain
\be
\frac{g^2 \tilde{\chi}_s^{(0)}(i \omega=0)}{M \Omega^2} \sim \frac{E_p}{\Delta_K} > 1
\ee
for physically relevant cases.

However, regarding the result Eq.~(\ref{phononprop1})  for which the bose fluctuations taken into account,
the ratio becomes
\be
\label{ratio1}
\frac{g^2 \tilde{\chi}_c^{(0)}(i \omega=0)}{M \Omega^2} \sim \frac{E_p \Delta_K}{\Delta_0^2} \ll 1,
\ee
which agrees with the results by NRG.
Thus, unless the electron-phonon interaction is extremely strong,
 the self-energy correction to the phonon propagator coming from the electronic sector
is small. But if the electron-phonon interaction becomes so strong that the ratio  Eq.~(\ref{ratio1}) becomes order of
unity, the saddle point we have started with won't be valid any more
and  new saddle points will emerge.  The results by NRG studies indeed demonstrate that
new features emerge in the strong coupling regime of electron-phonon interaction. Unfortunately our approximation scheme
does not apply in that regime.
In any case we conclude that
 to obtain the physically correct result it is absolutely essential to include the  $1/N$ bose fluctuations $X$.

The form of the phonon propagator Eq.~(\ref{phononprop1}) suggests that the phonon frequency softens.
The renormalized phonon frequency is given by
\be
\label{soft1}
\Omega_{ren} \sim \Omega \Big(1- \frac{\pi}{2}\,\frac{E_p \Delta}{\Delta_0^2} \Big).
\ee
The result Eq.~(\ref{soft1}) can be compared with that derived in the \textit{weak}  
(electron-electron and electron-phonon)
coupling random phase approximation
type perturbative calculations:\cite{jpc}
\be
\label{soft2}
\Omega_{ren} \sim \Omega \Big(1- \frac{E_p}{\pi \Delta_0} \Big).
\ee
The result Eq.~(\ref{soft1}) should be also compared with that derived in the semiclassical approximation
for the large but \textit{finite}-$U$ and \textit{symmetric} AH model:\cite{ours}
\be
\label{soft3}
\Omega_{ren} \sim \Omega \Big(1- \frac{E_p \Delta_0}{U^2} \Big).
\ee
The results Eqs.~(\ref{soft1},\ref{soft3}) indicate that the strong Coulomb repulsion suppresses
the softening of phonon frequency but the detailed form of the softening depends crucially on the details of
Hamiltonian such as the symmetric or asymmetric nature of charge fluctuations of Anderson model.
 %%%%%%%%%%%%%%%%%%%%%%%%%%%%%%%%%%%%%%%%%%%%%%%%%%%%%%%%%%
\section{Summary and Concluding Remarks}
We have studied the AH model with the infinite local Coulomb repulsion.
The limit of the infinite local Coulomb repulsion eliminates the doubly occupied states in the Hilbert space of impurity.
The slave boson method is very effective in treating the problems with the constrained Hilbert space,
 where the strong correlation
is actually built in the constraint itself. 
In general the slave boson method is more reliable in the low temperature regime where the thermal
fluctuations are small.
The $1/N$ expansion scheme is employed to control approximations systematically and it is found that the
electron-phonon interaction can be naturally fit into the scheme.  Remarkably the nonperturbative physics of the low temperature Kondo fixed point is captured
in the leading approximation.
We have used the saddle point approximation supplemented with the Gaussian fluctuations.
In fact, the Gaussian fluctuations play a \textit{very} crucial role in determining the physical
properties of impurity and phonon. The spin susceptibility is not renormalized up to this approximation. 
The Kondo temperature is found to increase by the elctron-phonon interaction but by a rather small amount.
The same is true of the charge susceptibility and phonon propagator.
This is because the quantum fluctuations of phonon which couple to the charge fluctuations of the impurity
electron are suppressed by the very strong local electron correlation. But the \textit{interplay} between the strong
electron correlation and the electron-phonon interaction is clearly displayed.

We could also have calculated the electron spectral function but  all the important features such as the increase
of the Kondo temperatures and the pronounced phonon effect near the phonon freqency are essentially identical
with those of the charge susceptibility.

The AH model can be naturally generalized to the impurity models with the orbital degrees of freedom.
As a matter of fact, this is much more realistic model than the AH model since the electron correlation effects
 are more prominent in the
$d$ and $f$ electron systems where the orbital degrees of freedom are active.
For these cases the simple Holstein phonon should  be also generalized to incorporate 
 Jahn-Teller type phonons.\cite{orbital}
Slave boson approach can be extended to the orbitally active cases, and currently
the progresses in this direction are being
made by the authors.
%%%%%%%%%%%%%%%%%%%%%%%%%%%%%%%%%%%%%%%%%%%%%%%%%%%%%%%%%%%%%%%
\begin{acknowledgments}
H.C.L  was  supported by the Korea Science and Engineering
through the Center for Strongly Correlated Materials Research (CSCMR) and through
Grant No. R01-1999-000-00031-0, and was also supported by the
Sogang University Research Grants in 2003.
H.Y.C was supported by the by the Korea Science and Engineering
Foundation (KOSEF) through the grant No. R01-1999-000-00031-0 and the
Ministry of Education through Brain Korea 21 SNU-SKKU Program.
\end{acknowledgments}
%%%%%%%%%%%%%%%%%%%%%%%%%%%%%%%%%%%%%%%%%%%%%%%%%%%%%%%%%%%%%%%%

\appendix
\section{Illustration of the basic idea of $1/N$ expansion}
\label{1/N}
Let us briefly illustrate the basic idea of $1/N$ expansion in the case of one-dimensional integral.
Consider an integral $I$ of the following form defined on a certain domain $D$:
\be
I = \int_D e^{-N g(x)},\;\; N \gg 1.
\ee
If a minimum of $g(x)$ is attained at $x_0 \in D$, then the function can be expanded
\be
g(x) \sim g(x_0) + \frac{1}{2} g^{(2)}(x_0) (x-x_0)^2+ \sum_{n \ge 3} \frac{g^{(n)}(x_0)}{n!}(x-x_0)^n,
\ee
where $g^{(n)}$ denotes the $n$-th derivative and $g^{(2)}(x_0) > 0$.
Defining a \textit{rescaled} variable $z$
\be
z = \frac{1}{\sqrt{N}}\,\frac{1}{\sqrt{g^{(2)}(x_0)}}\,(x-x_0)
\ee
the integral $I$ can be approximated by
\ba
\label{nintegral}
I &\sim& e^{-N g (x_0)-\ln [ N g^{(2)}(x_0)]/2}\,\int dz \,e^{-z^2/2-c_1 z^3/\sqrt{N}+\cdots} \nonumber \\
& \sim & e^{-N g (x_0)-\ln [ N g^{(2)}(x_0)]/2}\,\Big(\sqrt{2\pi}+\frac{\textrm{const.}}{\sqrt{N}}+\cdots\Big)
\ea
The integration over $z$ can be done systematically as a power series expansion in $1/\sqrt{N}$.
In this paper we  ignore the corrections of order $1/\sqrt{N}$ and higher in the bracket of
Eq.~(\ref{nintegral}), but
they can be computed in a controllable way if desired.
%%%%%%%%%%%%%%%%%%%%%%%%%%%%%%%%%%%%%%%%%%%%%%%%%%%%%%%%%%%%%%%%%%%%%%%%%%%%%
\section{Details of the calculations of  $1/N$ corrections}
\label{append:1/N}
First let us specify the measure for the functional integral in the RN gauge. This is discussed in
Sec. III B of Ref. \onlinecite{coleman}, and  here we give an equivalent but technically slightly different treatment.
The measure of boson functional integral is given by
\be
D[r,\theta]=\prod_\tau\, \frac{d \big(r^2(\tau)\,\theta(\tau) \big)}{2\pi}.
\ee
The fermion measure is invariant under the RN gauge transformation.
\be
D[s,s^\dag]=D[z,z^{\dag}].
\ee
The angular variable $\theta$ in the sector with the winding number $m$ can be expanded \cite{coleman}
\be
\theta^{(m)}(\tau)=\frac{1}{\sqrt{\beta}}\,\sum_{n}\,\theta_n\,e^{-i \nu_n \tau} +
2\pi m \frac{\tau}{\beta},
\ee
where $\nu_n$ is the bosonic Matsubara frequency given by $\nu_n=2 \pi n /\beta$.
This expansion automatically satisfies the winding number condition
without further conditions imposed on the coefficients $\theta_n$.
The zero mode term  $\theta_0$ should be factored out since our action does not depend on it.
For the variation which involves only the zero mode
we have $d \theta^{(m)}_0= 1/\sqrt{\beta} d \theta_0$.  Factoring out $\int d \theta^{(m)}_0$ we obtain
\be
D[\theta]=\sum_m\,\sqrt{\beta}\,\prod_{n \neq 0}\,  \frac{ d \theta_{n}}{2\pi}.
\ee
Once the zero mode term is factored out both $\theta(\tau)$ and $d \theta(\tau)/d\tau$ are characterized by the same set
of coefficients $\{ \theta_{n \neq 0} \}$ and an integer $m$.
Explicitly
$$ \frac{d \theta}{d \tau}=2\pi m /\beta + \frac{1}{\sqrt{\beta}}\,\sum_{n \neq 0}\,(-i \nu_n)\,\theta_n \,e^{-i \nu_n \tau}.$$
From the above expression we find that  the constant part  of  $\frac{d \theta}{d \tau}$  is constrained to take a value
which is an integral multiple of $2 \pi /\beta$. However,
as can be seen in Eq.~(\ref{action:rn3}) $\frac{d \theta}{d \tau}$ always appears in combination with $\Omega_0$. Since $\Omega_0$
is a part of Lagrange multiplier implementing delta function constraint it is non-compact, in other words,
 it can take an
arbitrary real value. Now the explicit summation over the winding number is unnecessary and we need only to consider a bose field
\be
\Theta= \frac{d \theta}{d \tau}+ \Omega_0,
\ee
which has an ordinary mode expasion without any constraint. Effectively we can replace $D[\theta]$ with $D[\Theta]$.

If the fluctuations beyond the Gaussian fluctuation of the order $1/N$ can be ignored, namely the corrections of
order $1/N^{3/2}$ and higher can be ignored, the functional measure for the $r$ integration can be approximated by
\be
d r^2 (\tau)= 2 r(\tau)  d r (\tau)  \sim 2 r_{sa} d \delta r (\tau).
\ee
The higher order corrections can be important at high temperature or energy.
We also note that due to the periodicity $r(0)=r(\beta)$, $\int_0^\beta d \tau r \frac{\partial r}{\partial \tau}=0$.
Thus, at low energy the  measure for the boson functional integral  becomes
\be
D[r, \theta]  \sim \prod_\tau  r_{sa}  d \delta r (\tau)   d \Theta (\tau).
\ee
This is of a simple Cartesian product form,
so that we don't have to deal with the interactions which might be generated
by the nontrivial Jacobians of  functional integral measure.

Now we turn to the details of the calculation of Eq.~(\ref{logdet}). Let us rewrite Eq.~(\ref{logdet})
as follows:
\ba
\label{eq:expansion}
\delta S&=&-\sum_m\,{\rm Tr}\ln[(-G_0^{-1})(1-G_0  M)] \nonumber \\
&=&-\sum_m \, {\rm Tr} \ln[-G_0^{-1}]-\sum_m\,{\rm Tr} \ln (1-G_0  M) \nonumber \\
&\sim&-\sum_m \, {\rm Tr} \ln[-G_0^{-1}]+\sum_m\,{\rm Tr} \big[ G_0  M \big ] \nonumber \\
&+&\frac{1}{2}\,\sum_m\,{\rm Tr} \big[G_0  M G_0  M \big].
\ea
In the absence of the source field $\phi_m$ the sum over $m$ would just give an overall
factor $N$. The first term of the last line of Eq.~(\ref{eq:expansion}) is relevant for the mean-field approximation.
The second term $\sum_m\,{\rm Tr}\ln [G_0  M]$ contains a few parts.
Among them $g Q$ part is cancelled by
the very definition of electron-phonon coupling, and
the part linear in $\delta r$ is cancelled by the saddle point condition.
The remaining nontrivial part is given by
\ba
\label{linear}
& &\sum_m\,{\rm Tr}\ln \big [G_0  M \big ] \to
\sum_m\,\int d \tau\, \phi_m G_0 \nonumber \\
&+&\frac{N}{2}\, \sum_{i \omega} \Big[ K_{rr}^{(0)}+K_{rr}^{(2)} \Big ]\,\delta r(i \omega)\,\delta r(-i \omega),
\ea
where $K_{rr}^{(0)}$ and $K_{rr}^{(2)}$ are given by
\ba
K_{rr}^{(0)}&=&2  T \sum_{i \epsilon} G_0(i\epsilon) \Sigma_0(i\epsilon), \\
K_{rr}^{(2)}(i\omega)&=&2  T \sum_{i \epsilon} G_0(i\epsilon)
\big[ \Sigma_0(i\epsilon+i\omega)-\Sigma_0(i\epsilon) \big].
\ea

The third term $\frac{1}{2}\,\sum_m\,{\rm Tr} \ln [G_0  M G_0  M ]$ can be re-expressed as follows:
\ba
\label{quadratic}
& &\frac{1}{2}\,\sum_m\,{\rm Tr} \ln [G_0  M G_0  M ]=
\frac{N}{2}\,\sum_{i \omega}\,K_{rr}^{(3)}\,\delta r(i \omega) \delta r(-i \omega)  \nonumber \\
& &\half \sum_{m} \sum_{i \omega} \,K_{\theta \theta}^{(1)}\,r_{sa}^2 ( i \Theta + \frac{g}{\sqrt{N}} Q +\phi_m)(i\omega) \nonumber \\
& &\times ( i \Theta + \frac{g}{\sqrt{N}} Q +\phi_m)(-i\omega) \nonumber  \\
& &+\half \sum_m \sum_{i\omega}\,2 K_{r \theta}^{(1)}\,\delta r (i\omega) \,
 r_{sa} ( i \Theta + \frac{g}{\sqrt{N}} Q +\phi_m)(-i\omega).
\ea
where
\ba
K_{rr}^{(3)}(i\omega)&=&r_{sa}^2 T \sum_{i \epsilon}\,G_0(i \epsilon) G_0(i \epsilon  +i \omega) \nonumber \\
&\times& \Big( \Sigma_0(i \epsilon)+ \Sigma_0(i \epsilon+i \omega) \Big)^2, \nonumber \\
K_{\theta \theta}^{(1)}(i\omega)&=&\frac{1}{r_{sa}^2} T \sum_{i\epsilon} G_0(i \epsilon) G_0(i \epsilon  +i \omega),\nonumber \\
 K_{r \theta}^{(1)}&=&
 T \sum_{i \epsilon}\,G_0(i \epsilon) G_0(i \epsilon  +i \omega) \nonumber \\
&\times& \Big( \Sigma_0(i \epsilon)+ \Sigma_0(i \epsilon+i \omega) \Big).
\ea
%%%%%%%%%%%%%%%%%%%%%%%%%%%%%%%%%%%%%%%%%%%%%%%%%%%%%%%%%%%%%%%%%%%%%%%%%%%
\section{The explicit forms of the polarization functions $K^{(i)}(i \omega)$ at zero temperature}
\label{polarization}
The polarization functions $K(i \omega)$ have been calculated in Appendix A of Ref. \onlinecite{coleman}
in terms of the digamma function at finite temperature.
We will instead calculate the polarization functions at zero temperature  in a closed form.
The integrations are completely elementary, and only the results and the low frequency asymptotic behaviors
will be displayed.
First we note  that only the real part of $K_{rr}^{(2)}(i\omega)$  which is an even function
of frequency contributes to the effective action.
\be
{\rm Re} K_{rr}^{(2)}(i\omega)=\frac{\Delta_0}{\pi}\,
\ln
\Big [ \frac{(|\omega| + \Delta)^2+\tilde{\epsilon}_f^2}{
   \Delta^2 + \tilde{\epsilon}_f^2} \Big ].
\ee

\be
K_{rr}^{(3)}(i\omega)= \frac{2 \Delta_0 \Delta}{ \pi |\omega|}\,\ln
\Big [ \frac{(|\omega| + \Delta)^2+\tilde{\epsilon}_f^2}{
   \Delta^2 + \tilde{\epsilon}_f^2} \Big ].
\ee
The sum of the above two gives
\ba
K_{rr}^{(1)}(i\omega)& \equiv &{\rm Re} K_{rr}^{(2)}+K_{rr}^{(3)} \nonumber \\
&=&+ \frac{\Delta_0}{\pi}\, \frac{ |\omega| + 2 \Delta}{|\omega|}\,
\ln \Big[ 1+ \frac{|\omega|(|\omega|+2 \Delta)}{\Delta_K^2} \Big].
\ea
And the rest are given by
\be
K_{r \theta}^{(1)}(i\omega)= \frac{2 \Delta_0}{\pi |\omega|}\,\tan^{-1}\,
\left [ \frac{ \tilde{\epsilon}_f |\omega|}{\Delta^2 + \tilde{\epsilon}_f^2 + |\omega| \Delta} \right ].
\ee
\be
K_{\theta \theta}^{(1)}(i\omega)=-\frac{\Delta_0}{\pi|\omega| (|\omega|+2 \Delta)}\,
\ln \Big[1+ \frac{|\omega|(2  \Delta  +|\omega|)}{\Delta_K^2} \Big].
\ee
The low (imaginary) frequency asymptotic behaviors are given by
\be
K_{r \theta}^{(1)}(i\omega )  \sim \frac{4 \Delta_0}{\pi}
\,\Big[\frac{\Delta^2}{\Delta_K^2}+\frac{\Delta |\omega| \tilde{\epsilon}_f^2}{\Delta_K^4}\Big].
\ee
\be
K_{r \theta}^{(1)}(i\omega )  \sim \frac{2 \Delta_0}{\pi}\,\frac{\tilde{\epsilon}_f}{
\Delta^2+\tilde{\epsilon}_f^2}.
\ee
\be
 K_{\theta \theta}^{(1)}(i\omega) \sim - \frac{1}{\pi r_{sa}^2}\,\frac{\Delta}{\Delta^2+\tilde{\epsilon}_f^2}
 + \frac{1}{\pi r_{sa}^2}\,\frac{|\omega| \Delta^2}{(\Delta^2+\tilde{\epsilon}_f^2)^2}.
\ee

%%%%%%%%%%%%%%%%%%%%%%%%%%%%%%%%%%%%%%%%%%%%%%%%%%%%%%%%%%%

%%%%%%%%%%%%%%%%%%%%%%%%%%%%%%%%%%%%%%%%%%%%%%%%%%%%%%%%%%%%

\begin{thebibliography}{}
\bibitem{comment} We consider  the one-band and spin 1/2 case only, 
so that the underscreened and overscreened Kondo states
are not relevant to our discussions.
\bibitem{notation} We use the notation $s_\sigma$ for the constrained electron operator, while
$f_\sigma$ is used in Coleman's paper [\onlinecite{coleman}].
\bibitem{note1} In calculation of the physical quantities for $N=2$ case, 
the rescaling $ V_k  \to \sqrt{2} V_k$ is necessary.
\end{thebibliography}
\end{document}